\documentclass[11pt,superscriptaddress,aps,prd,preprint]{revtex4}

\everymath{\displaystyle}
\usepackage{graphicx}
\usepackage{amsmath,amssymb,amsfonts,amstext,latexsym,stmaryrd,amscd,txfonts,pxfonts}

\begin{document}

\title{Non-linear Plane Gravitational Waves as Space-time Defects}

\author{F. L. Carneiro}\email[]{fernandolessa45@gmail.com}
\affiliation{Instituto de F\'{\i}sica, 
Universidade de Bras\'{\i}lia\\
70.919-970 Bras\'{\i}lia DF, Brazil}

\author{S. C. Ulhoa}\email[]{sc.ulhoa@gmail.com}
\affiliation{International Center of Physics, Instituto de F\'isica, Universidade de Bras\'ilia, 70910-900, Bras\'ilia, DF, Brazil} \affiliation{Canadian Quantum Research Center,\\ 
204-3002 32 Ave Vernon, BC V1T 2L7  Canada} 

\author{J. W. Maluf}
\email{jwmaluf@gmail.com} \affiliation{Instituto de F\'{\i}sica, 
Universidade de Bras\'{\i}lia\\
70.919-970 Bras\'{\i}lia DF, Brazil}

\author{J. F. da Rocha-Neto}\email[]{rocha@fis.unb.br}
\affiliation{Instituto de F\'{\i}sica, 
Universidade de Bras\'{\i}lia\\
70.919-970 Bras\'{\i}lia DF, Brazil}

\begin{abstract}
We consider non-linear plane gravitational waves as propagating space-time
defects, and construct the Burgers vector of the waves. In the context of
classical continuum systems, the Burgers vector is a measure of the 
deformation of the medium, and at a microscopic (atomic) scale, it is a
naturally quantized object. One purpose of the present article is ultimately
to probe an alternative way on how to quantize plane gravitational waves. 
\end{abstract}
\maketitle

\date{\today}
\section{Introduction} \label{sec.1}

Non-linear gravitational waves constitute a class of exact solutions of Einstein's field equations of general relativity. Several of these exact solutions form
a subset of solutions known as plane gravitational waves, or 
pp-waves (plane-fronted gravitational waves). These waves, in
turn, are characterized by a local (in space and in time) deformation of an
otherwise flat space-time. For this reason, it is possible to think of
non-linear plane gravitational waves as propagating space-time defects,
since the latter are also locally flat.

Non-linear plane gravitational waves and space-time defects share many 
features. Both field configurations (i) are established over a flat
space-time background, (ii) induce a local deformation in the background
geometry, (iii) may have an axial symmetry (along the $z$ axis, for
instance), (iv) may have a singularity along an axis (the $z$ axis, for
instance). Therefore, it is possible to define and evaluate the Burgers
vector for non-linear plane gravitational waves. The Burgers vector in a
crystalline lattice or inside a metal determines the nature of the defect.

Metals may be deformed both elastically and plastically. Elastic 
deformations take place at low external stresses, and are reversible,
whereas plastic deformations are irreversible. The latter deformations are
described by moving, dynamical dislocations, which imprint a permanent 
deformation in the metal. The two ordinary types of dislocations in a metal
are the screw and edge dislocations. A screw dislocation occurs when a 
half plane moves, or slips, relatively to one adjacent half plane (consider
two infinite, adjacent and parallel planes, the upper and lower planes; one
upper half plane remains attached to the adjacent lower half plane, whereas
the other upper half plane 
slips with respect to the adjacent lower half plane), and an
edge dislocation is characterized by a missing half plane in an otherwise
perfect lattice. We refer the reader to Chapter 5 of Ref. \cite{Smallman},
where a clear explanation of these defects is presented. The Burgers vector
is constructed by first establishing a Burgers circuit, which is a 
circuit around the defect. The idea is to first consider a closed,
regular circuit in a perfect crystalline lattice. If the lattice is 
deformed by one type of dislocation, the circuit established in the perfect
lattice fails to close in the deformed medium, 
if the circuit encompasses the defect.
The vector that must be added in order to close the 
circuit, in the presence of a defect, is precisely the Burgers vector.
At the atomic scale in a physical lattice, the Burgers vector is quantized,
i.e., it is a multiple of a minimum atomic distance.
In the context of the 4-dimensional space-time geometry, space-time (or 
topological) defects have already been investigated in some depth, see 
for instance Refs. \cite{Katanaev,Holz,Puntigam,Todd,Bilby}.

Non-linear gravitational waves are relatively simple solutions of
Einstein's equations, as we learn from the well known review by 
Ehlers and Kundt \cite{Ehlers}. These waves have been recently reconsidered in 
connection with the memory effect \cite{ZDGH1,ZDGH2,ZDGH3,ZDGH4}. 
The memory effect is the permanent displacement of massive particles and
ordinary objects of a physical system 
caused by the passage of a non-linear gravitational wave (although 
the memory effect is also considered in the context of linearised 
gravitational waves). In particular, the dynamical state of the massive 
particles is different before and after the passage of the wave 
\cite{Artigo1,Artigo2,Artigo3,Artigo4}, in view of the velocity
memory effect. The
``permanent displacement" mentioned above may be understood as a plastic
deformation of the physical medium, that is constituted by massive 
particles, and in this sense propagating defects in metals (and crystalline
lattices) and non-linear gravitational waves share relevant features.
On the other hand, linearised gravitational waves may be understood as 
elastic deformations of the space-time.

In this article we will consider several relevant circuits in the space-time
of pp-waves and construct the corresponding Burgers vector. The graphical 
distribution of Burgers vectors in the three dimensional space allows an
alternative understanding and characterization of these waves. We will also 
evaluate the gravitational pressure that the pp-waves exert on certain 
surfaces, that allows to obtain the gravitational force applied
on idealized particles. The evaluation of the gravitational pressure is 
carried out with the definitions established in the teleparallel equivalent 
of general relativity. Under certain approximations, the resulting 
expressions of the gravitational pressure are quite simple and interesting.

This article is divided as follows. In section \ref{sec.2} the  Teleparallelism Equivalent to General Relativity is described. In section \ref{sec.3} the generalized pp-waves are introduced. The strain tensor and the Burgers vector are evaluated also in section III. In section \ref{sec.5}, the gravitational pressure and the force exerted by the wave are calculated. Finally in the last section the conclusions are presented. We use geometrical unities system where $G=c=1.$
\section{Teleparallelism Equivalent to General Relativity (TEGR)} \label{sec.2}

In this section we briefly introduce the ideas of Teleparallelism Equivalent to General Relativity (TEGR) along the lines of reference \cite{maluf2013teleparallel}. In this approach, the gravitational field is represented in terms of the dynamic tetrad field $e^{a}\,_{\mu} $, but at the same time it establishes the reference system by choosing the six additional components when compared to the metric tensor. The geometric framework of TEGR is such that absolute parallelism is a fundamental attribute of space-time. This condition is determined by the Weitzenb\"ock connection
$$\Gamma_{\mu\lambda\nu}=e^{a}\,_{\mu}\partial_{\lambda}e_{a\nu}\,$$
which has a vanishing curvature and a torsion tensor defined by
\begin{equation}
T^{a}\,_{\lambda\nu}=\partial_{\lambda} e^{a}\,_{\nu}-\partial_{\nu}
e^{a}\,_{\lambda}\,. \label{4}
\end{equation}
The Weitzenb\"ock connection is related to the Christoffel's symbols, ${}^0\Gamma_{\mu \lambda\nu}$, identically by
\begin{equation}
\Gamma_{\mu \lambda\nu}= {}^0\Gamma_{\mu \lambda\nu}+ K_{\mu
\lambda\nu}\,, \label{2}
\end{equation}
where $K_{\mu \lambda\nu}$ is the contortion tensor, and is given by
\begin{eqnarray}
K_{\mu\lambda\nu}&=&\frac{1}{2}(T_{\lambda\mu\nu}+T_{\nu\lambda\mu}+T_{\mu\lambda\nu})\,,\label{3}
\end{eqnarray}
with $T_{\mu \lambda\nu}=e_{a\mu}T^{a}\,_{\lambda\nu}$. The expression (\ref{2}) induces a direct relationship between Ricci scalar and a quadratic combination of torsions. It reads
\begin{equation}
eR(e)\equiv -e({1\over 4}T^{abc}T_{abc}+{1\over
2}T^{abc}T_{bac}-T^aT_a)+2\partial_\mu(eT^\mu)\,.\label{eq5}
\end{equation}
It should be noted that the left-hand side of the above expression is the Hilbert-Einstein Lagrangian. Thus the TEGR Lagrangian density is given by
\begin{eqnarray}
\mathfrak{L}(e_{a\mu})&=& -\kappa\,e\,({1\over 4}T^{abc}T_{abc}+
{1\over 2} T^{abc}T_{bac} -T^aT_a) -\mathfrak{L}_M\nonumber \\
&\equiv&-\kappa\,e \Sigma^{abc}T_{abc} -\mathfrak{L}_M\;, \label{6}
\end{eqnarray}
where $\kappa=1/(16 \pi)$, $\mathfrak{L}_M$ is the Lagrangian density of matter fields and $\Sigma^{abc}$ is defined as
\begin{equation}
\Sigma^{abc}={1\over 4} (T^{abc}+T^{bac}-T^{cab}) +{1\over 2}(
\eta^{ac}T^b-\eta^{ab}T^c)\;, \label{7}
\end{equation}
with $T^a=e^a\,_\mu T^\mu=e^a\,_\mu T^{\nu}\,_{\nu}\,^{\mu}$. Hence the field equations that are equivalent to Einstein's equations read
\begin{equation}
\partial_\nu\left(e\Sigma^{a\lambda\nu}\right)={1\over {4\kappa}}
e\, e^a\,_\mu( t^{\lambda \mu} + T^{\lambda \mu})\;, \label{10}
\end{equation}
where
\begin{equation}
t^{\lambda \mu}=\kappa\left[4\,\Sigma^{bc\lambda}T_{bc}\,^\mu- g^{\lambda
\mu}\, \Sigma^{abc}T_{abc}\right]\,, \label{11}
\end{equation}
is the gravitational energy-momentum tensor. Such an expression goes one step further towards the solution of the longstanding problem of gravitational energy.

Due to the anti-symmetric feature of $\Sigma^{a\lambda\nu}$, it is possible to obtain
\begin{equation}
\partial_\lambda\partial_\nu\left(e\Sigma^{a\lambda\nu}\right)\equiv0\,.\label{12}
\end{equation}
Then the energy-momentum vector is
\begin{equation}
P^a = \int_V d^3x \,e\,e^a\,_\mu(t^{0\mu}+ T^{0\mu})\,, \label{14}
\end{equation}
this can be equivalently expressed as
\begin{equation}
P^a =4k\, \int_V d^3x \,\partial_\nu\left(e\,\Sigma^{a0\nu}\right)\,. \label{14.1}
\end{equation}
It should be noted that the energy-momentum vector is invariant under coordinate transformations of the 3-dimensional space, and under time reparametrizations. On the other hand, it transforms as a vector under SO(3,1) symmetry. 

From equations (\ref{10}) and (\ref{12}) one obtains
$$
{d\over {dt}} \int_V d^3x\,e\,e^a\,_\mu (t^{0\mu} +T^{0\mu})
=-\oint_S dS_j\, \left[e\,e^a\,_\mu (t^{j\mu} +T^{j\mu})\right] \,,
$$
which indicates an energy-momentum flux, where the surface $S$ delimits the volume $V$ . The gravitational energy-momentum flux is defined as
\begin{equation}
\Phi^a_g=\oint_S dS_j\,
\, (e\,e^a\,_\mu t^{j\mu})\,,
\label{15}
\end{equation}
and the energy-momentum flux of matter fields as
\begin{equation}
\Phi^a_m=\oint_S dS_j\,
\,( e\,e^a\,_\mu T^{j\mu})\,.
\label{16}
\end{equation}
Then, it is possible to write
\begin{eqnarray}
{{dP^a}\over {dt}}&=&-\left(\Phi^a_g+\Phi^a_m\right)\nonumber\\
&=&-4k\oint_S dS_j\,
\partial_\nu(e\Sigma^{a j\nu})\,.
\label{17}
\end{eqnarray}
The momentum flux is given by the spatial part of the above equation, thus
\begin{equation}
{{dP^{(i)}}\over {dt}}= -\oint_S dS_j\, \phi^{(i)j} \,,
\label{18}
\end{equation}
where
\begin{equation}
\phi^{(i)j}=4k\partial_\nu(e\Sigma^{(i)j\nu}) \,.
\label{19}
\end{equation}
Equation (\ref{18}) has the nature of force, therefore equation (\ref{19}) represents the pressure on a direction $(i)$ over a surface oriented towards $j$.

\section{The generalized pp-waves} \label{sec.3}

The line element of the pp waves can be described in double null coordinates $ u, v $ by the generalized line element
\begin{equation}\label{linele}
ds^{2}=H(u,x,y)du^{2}+dx^{2}+dy^{2}+2dudv-2a_{1}(u,x,y)dudx -2a_{2}(u,x,y)dudy\,.
\end{equation}
The surfaces $u=constant$ are flat and the wave propagates along the null direction $v$.
The functions $a_{1,2}(u,x,y)$ may be eliminated locally by an appropriate choice of coordinates, therefore they may be chosen as zero. However, some topological properties of the space-time may be lost when such a choice is made \cite{podolsky2014gyratonic}. Topological defects manifests as a global effect, thus it is worth considering the generalized form of the pp-waves metric and particularizing it latter as special cases.

The line element (\ref{linele}) may be written in Cartesian coordinates by means of the relations
\begin{equation}
u=\frac{z-t}{\sqrt{2}}\,,\qquad v=\frac{z+t}{\sqrt{2}}\,.
\end{equation}
The line element becomes
\begin{align}
ds^{2}&=\left(\frac{H}{2}-1\right)dt^{2}+dx^{2}+dy^{2}+\left(\frac{H}{2}+1\right)dz^{2}-Hdtdz\nonumber\\
&+\sqrt{2}a_{1}dtdx-\sqrt{2}a_{1}dxdz +\sqrt{2}a_{2}dtdy-\sqrt{2}a_{2}dydz\,.\label{eq3}
\end{align}
The regular pp-waves may be obtained by choosing $a_{1,2}(u,x,y)=0$, as mentioned above. 

A tetrad field adapted to a stationary observer and related to the line element above can be can be written as
\begin{equation}\label{tet}
e_{a\mu}=\left(
\begin{array}{cccc}
 -A& a_{1}/\sqrt{2}A &  a_{2}/\sqrt{2}A & -H/2A \\
 0 & 1 & 0 & 0 \\
 0 & 0 & 1 & 0 \\
 0 &  -a_{1}/\sqrt{2}A & - a_{2}/\sqrt{2}A & 1/A \\
\end{array}
\right)\,,
\end{equation}
where $a$ and $\mu$ denote lines and rows, respectively, and $A=\sqrt{1-H/2}$.

A particular class of pp-waves that can be described by the line element (\ref{linele}) are the gyratonic waves \cite{frolov2005gravitational,frolov2005gravitational2,PRD75}. Those waves represent the outer field of the gyratons, which are spinning particles moving at the speed of light. In order to describe these waves, we introduce cylindrical coordinates in the transverse plane of the propagation line by means of the standard relations
\begin{align}
x&=\rho\cos{\phi}\,,\nonumber\\
y&=\rho\sin{\phi}\,,\nonumber
\end{align}
together with the functions, 
\begin{align}
a_{1}&=-\frac{J}{\rho}\sin{\phi}\,,\\
a_{2}&=\frac{J}{\rho}\cos{\phi}\,.
\end{align} 
Then, the line element (\ref{eq3}) becomes
\begin{equation}
ds^{2}=\left(\frac{H}{2}-1\right)dt^{2}+d\rho^{2}+\rho^{2}d\phi^{2}+\sqrt{2}Jdtd\phi-\sqrt{2}Jdzd\phi+\left(1+\frac{H}{2}\right)dz^{2}-Hdtdz\,.
\end{equation}
The function $J$ is related to the spinning nature of the gyratons.
Similarly, the tetrad field adapted to a stationary observer for the above metric may be rewritten as
\begin{equation}\label{tetcil}
e'_{a\mu}=\left(
\begin{array}{cccc}
 -A& 0 & \frac{J}{\sqrt{2}A} & -H/2A \\
 0 & \cos (\phi ) & -\rho  \sin (\phi ) & 0 \\
 0 & \sin (\phi ) & \rho  \cos (\phi ) & 0 \\
 0 & 0 & -\frac{J}{\sqrt{2}A} & 1/A \\
\end{array}
\right)\,.
\end{equation}
It should be noted that for $J=0$ the regular pp-waves are recovered. From the field equations, the relation between $H$ and $J$ is given by
\begin{equation}\label{einsequa}
\nabla^{2}H=\frac{2}{\rho^{2}}\left(\partial_{u}\partial_{\phi}J\right)\,,
\end{equation}
where $\nabla^{2} \equiv \partial_{\rho}\partial_{\rho}+\frac{1}{\rho}\partial_{\rho}+\frac{1}{\rho^{2}}\partial_{\phi}\partial_{\phi}$ in cylindrical coordinates. For $J=J(u)\Rightarrow\partial_{\phi}J=0$
\begin{equation}\label{einseq}
\nabla^{2}H=0\,,
\end{equation}
then it is possible  to solve the equation (\ref{einsequa}) to obtain the following classes of solutions 
\begin{align}
H_{0}&=-C_{0}\ln{\left(\frac{x^{2}+y^{2}}{a^{2}}\right)}f(u)\,,\label{eq13}\\
H_{1+}&=-C_{1+}\left(x^{2}-y^{2}\right)f(u)\,,\label{eq14}\\
H_{1\times}&=-C_{1\times}\left(xy\right)f(u)\,,\label{eq15}\\
H_{2+}&=-C_{2+}\frac{x^{2}-y^{2}}{\left(x^{2}+y^{2}\right)^{2}}f(u)\,,\label{eq16}\\
H_{2\times}&=-C_{2\times}\frac{xy}{\left(x^{2}+y^{2}\right)^{2}}f(u)\,.\label{eq17}\end{align}
Here, the multiplicative factors have a proper dimension to leave $H$ dimensionless, for instance $C_{1+}$ and $C_{1\times}$ have dimension of inverse squared distance, $C_{2+}$ and $C_{2\times}$ have dimension of squared distance, while $C_0$ is dimensionless. The constant $a$ in (\ref{eq13}) delimits the validity region of the solution in vacuum, i.e., the radius of the source. These functions are not the only possible solutions, but they have a well establish physical meaning. The function $f(u)$ is arbitrary and establishes the form of the pulse, usually chosen as a Gaussian.
In the next subsections, effects such as deformations and distortions associated with generalized pp-waves will be analyzed.

\subsection{The strain tensor}

In solid mechanics, the deformation of materials is an important feature in understanding their properties. When an elastic deformation is present, the strain tensor quantifies the relative amount of change during deformation. In the case of plastic deformations such as dislocations, where Hooke's law does not apply everywhere, a dislocation core is constructed and outside this core the Hooke's law is applied. In a plastic deformation, the components of the strain tensor can be written as a function of the dislocation intensity, i.e., the Burgers vector.

We can import this concept into space-time.  Thus we introduce the strain tensor, understood as the difference between the geometries before and after a given event, for instance, the passage of a gravitational wave. Therefore the strain tensor is defined as \cite{deformation}
\begin{equation}
\varepsilon_{\mu\nu}\equiv \left(g_{\mu\nu}-\bar{g}_{\mu\nu}\right)
\end{equation}
where $\bar{g}_{\mu\nu}$ is the flat space-time metric in an arbitrary coordinate system and $g_{\mu\nu}$ is the metric tensor of the gravitational wave. Usually, in the case of metals, this tensor is built in three dimensions.  Here it is always possible to take the three-dimensional part of the deformation tensor for comparison. 

The strain tensor calculated from the metric (\ref{eq3}) is given by
\begin{equation}
\varepsilon_{\mu\nu}=\frac{1}{2}\left(
\begin{array}{cccc}
 H& -\sqrt{2}a_{1} &  -\sqrt{2}a_{2} & H \\
 -\sqrt{2}a_{1} & 0 & 0 & -\sqrt{2}a_{1} \\
 -\sqrt{2}a_{2}  & 0 & 0 & -\sqrt{2}a_{2} \\
 H & -\sqrt{2}a_{1} & -\sqrt{2}a_{2} & H \\
\end{array}
\right)\,.
\end{equation}
A deformation is a measurable effect, that is, it is not dependent on the chosen coordinate system.  Hence, it is necessary to project such a quantity on the Lorentz symmetry indices. We get then
\begin{equation}
\varepsilon^{ab}=e^{a\mu}e^{b\nu}\varepsilon_{\mu\nu}= \frac{1}{2}\left(
\begin{array}{cccc}
 H/A& -\sqrt{2}a_{1}/A &  -\sqrt{2}a_{2}/A & H/A \\
 -\sqrt{2}a_{1}/A & 0 & 0 & -\sqrt{2}a_{1}/A \\
 -\sqrt{2}a_{2}/A  & 0 & 0 & -\sqrt{2}a_{2}/A \\
 H/A &  -\sqrt{2}a_{1}/A & -\sqrt{2}a_{2}/A & H/A \\
\end{array}
\right)\,.
\end{equation}
As a consequence the 3D strain tensor of a gyratonic wave is
\begin{equation}\label{3dd}
\varepsilon^{(i)(j)}= \frac{1}{2}\left(
\begin{array}{ccc}
 0 & 0 & \frac{\sqrt{2}J}{\rho A}\sin{\phi} \\
 0 & 0 & -\frac{\sqrt{2}J}{\rho A}\cos{\phi} \\
\frac{\sqrt{2}J}{\rho A}\sin{\phi} & -\frac{\sqrt{2}J}{\rho A}\cos{\phi} & H/A \\
\end{array}
\right)\,,
\end{equation}
therefore it is possible to see that regular pp-waves are responsible for longitudinal deformations while gyratonics also cause transversal shearing.

\subsection{The Burgers vector} \label{sec.4}

In a spacetime with torsion, the Burgers vector is defined as

\begin{equation}
b^{a}=\frac{1}{2}\int_{S}T^{a}\,_{\mu\nu} dx^{\mu}\wedge dx^{\nu}\,,
\end{equation}
where $\wedge$ is the exterior product. This means that torsion is the dislocation superficial density. Due to the torsion symmetry and the properties of tetrad (\ref{tet}), the spatial components of the Burgers vector can be written as
$$b^{(i)}=\oint_{\mathcal{C}}{e^{(i)}\,_{j}dx^{j}} \,,$$ where $C$ is a path delimited by the surface $S$. It is worth noting that the result of this integral depends on the path taken.

In order to construct the Burgers circuit, a plane can be chosen. First, let us consider
a square with side $2L$ in the YZ plane, centered at an arbitrary point $(t_{0},x_{0},y_{0},z_{0})$. Thus
\begin{align}
b^{(i)}&=\int^{y_{0}+L}_{y_{0}-L}{e^{(i)}\,_{2}(t_{0},x_{0},y,z_{0}+L)dx^{2}}+\int^{z_{0}-L}_{z_{0}+L}{e^{(i)}\,_{3}(t_{0},x_{0},y_{0}-L,z)dx^{3}}\nonumber\\
&+\int^{y_{0}-L}_{y_{0}+L}{e^{(i)}\,_{2}(t_{0},x_{0},y,z_{0}-L)dx^{2}}+\int^{z_{0}+L}_{z_{0}-L}{e^{(i)}\,_{3}(t_{0},x_{0},y_{0}+L,z)dx^{3}}\,.
\end{align}
Then, the only non-vanishing component of the Burgers vector is 
$$b^{(3)}=-\frac{1}{\sqrt{2}}\int^{y_{0}+L}_{y_{0}-L}{\left.\frac{a_{2}}{A}\right|^{z=z_{0}+L}_{z=z_{0}-L}dy}+\int^{z_{0}+L}_{z_{0}-L}{\left.\frac{1}{A}\right|^{y=y_{0}+L}_{y=y_{0}-L}dz}\,.$$
The result for a regular pp-wave can be obtained by choosing $a_{1,2}=0$, yielding
\begin{equation}
b^{(3)}=\int^{z_{0}+L}_{z_{0}-L}{\left.\frac{1}{A}\right|^{y=y_{0}+L}_{y=y_{0}-L}dz}\,.
\end{equation}
We must note that the Burgers vector is calculated locally, therefore depends on the circuit shape and the location of its center. This means that we have a distribution of vectors in space, i.e., a vector field.
Each set of coordinates $(t_{0},x_{0},y_{0},z_{0})$ will have its own Burgers vector within its neighborhood.
In the figures below we show the form of this distribution for some choices of $H$, all for a regular pp-wave.
All figures were obtained for $2L=0.002$ and $f(u)=e^{-u^{2}}$. In all of them below, the side bar indicates a color scale for the vector modulus.

\begin{figure}[htbp]
	\centering
		\includegraphics[width=0.9\textwidth]{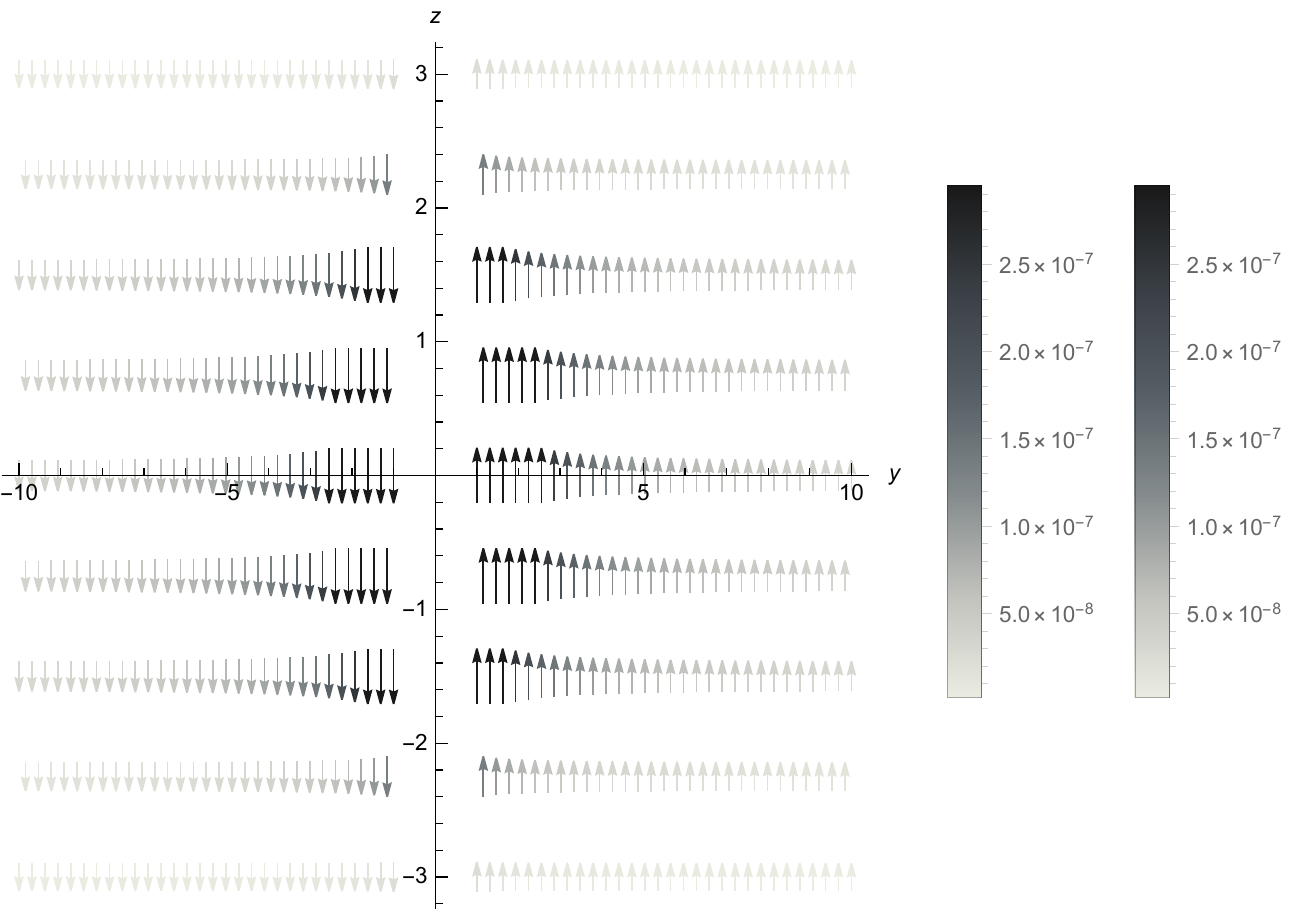}
	\caption{Burgers vector for $H_{0}$ with $t=x=0$ and $C_{0}=1$. The points were distributed with $y_{0}$ ranging from $-10$ to $-1$ and from $+1$ to $+10$, in steps of $0.05$; and $z_{0}$ ranging from $-3$ to $+3$, in steps of $0.1$. The region around the propagation axis was excluded from integration.}
\end{figure}
\begin{figure}[htbp]
	\centering
		\includegraphics[width=0.9\textwidth]{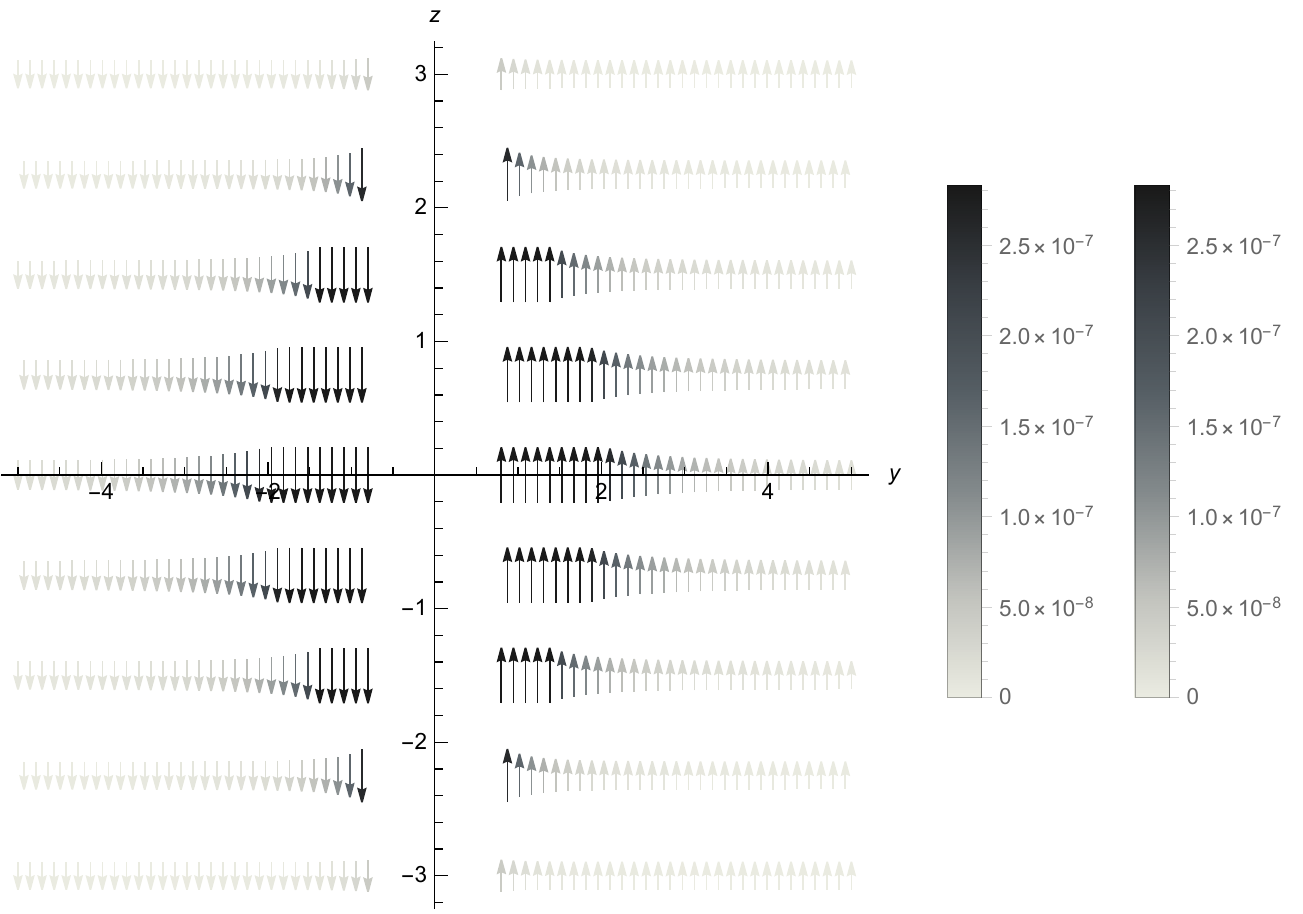}
	\caption{Burgers vector for $H_{2+}$ with $t=x=0$ and $C_{2+}=1$. The points were distributed with $y_{0}$ ranging from $-5$ to $-0.8$ and from $+0.8$ to $+5$, in steps of $0.05$; and $z_{0}$ ranging from $-3$ to $+3$, in steps of $0.1$. The region around the propagation axis was excluded from integration. The bars on the right side indicate the modulus of the Burgers vector on both sides.}
\end{figure}
\begin{figure}[htbp]
	\centering
		\includegraphics[width=0.9\textwidth]{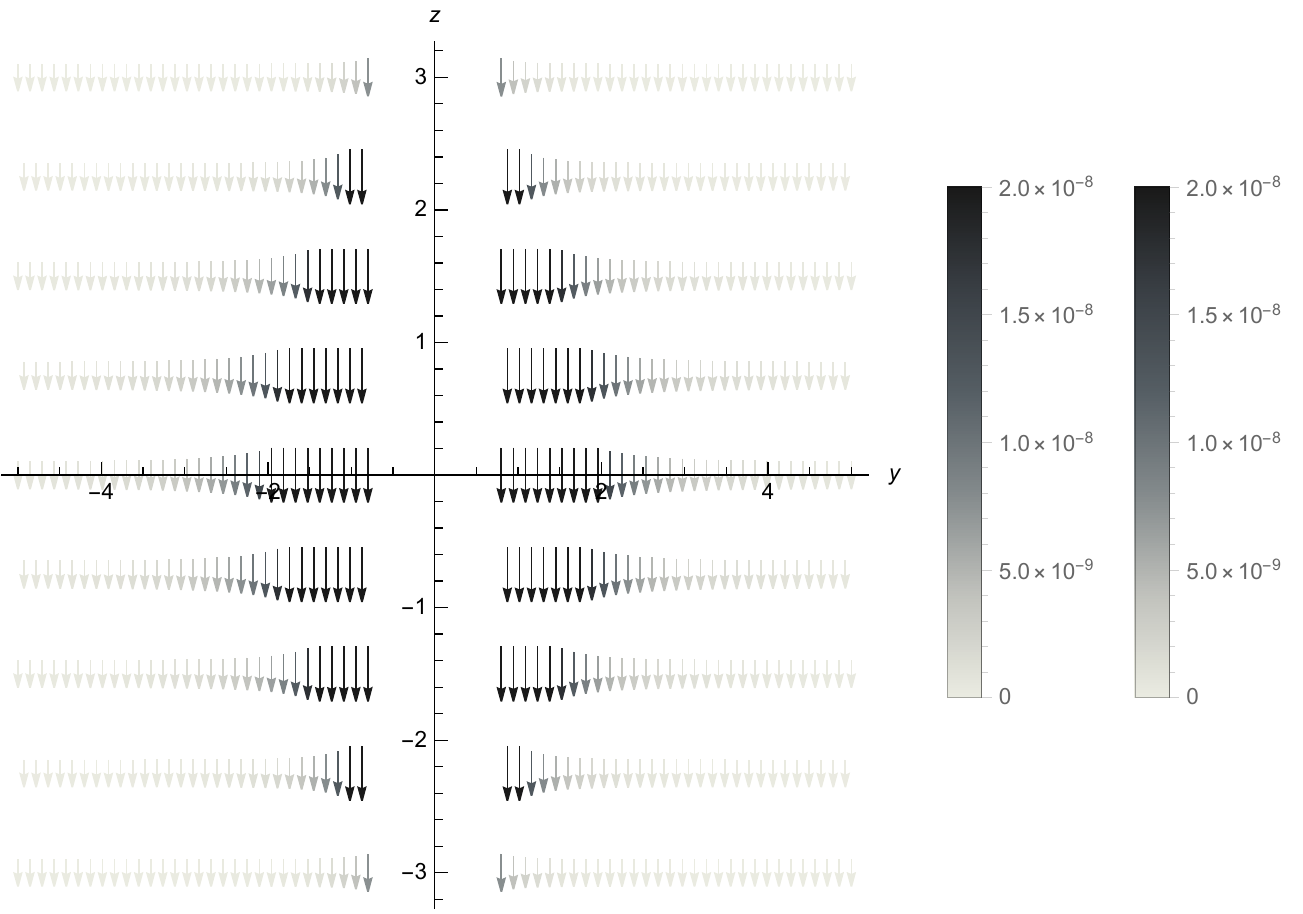}
	\caption{Burgers vector for $H_{2\times}$ with $t=0$, $x=0.1$ and $C_{2+}=1$. The points were distributed with $y_{0}$ ranging from $-5$ to $-0.8$ and from $+0.8$ to $+5$, in steps of $0.05$; and $z_{0}$ ranging from $-3$ to $+3$, in steps of $0.1$. The region around the propagation axis was excluded from integration. The bars on the right side indicate the modulus of the Burgers vector on both sides.}
\end{figure}

\begin{figure}[htbp]
	\centering
		\includegraphics[width=0.9\textwidth]{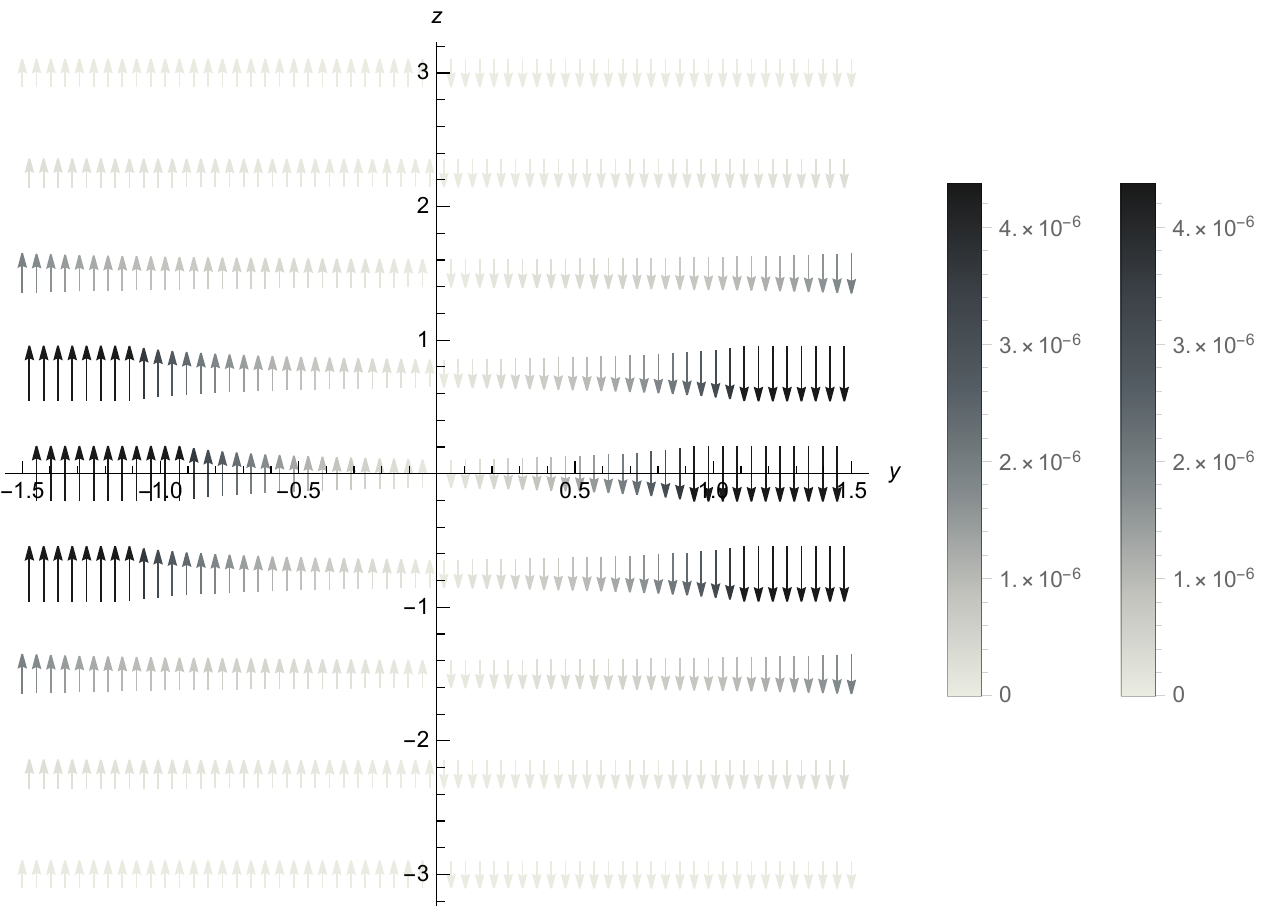}
	\caption{Burgers vector for $H_{1+}$ with $t=x=0$ and $C_{2+}=a=1$. The points were distributed with $y_{0}$ ranging from $-1.5$ to $1.5$, in steps of $0.05$; and $z_{0}$ ranging from $-3$ to $+3$, in steps of $0.1$. The bars on the right side indicate the modulus of the Burgers vector on both sides.}
\end{figure}

If we choose a similar circuit in the plane XZ, we obtain
$$b^{(3)}=-\frac{1}{\sqrt{2}}\int^{L}_{-L}{\left.\frac{a_{1}}{A}\right|^{z=L}_{z=-L}dx}+\int^{L}_{-L}{\left.\frac{1}{A}\right|^{x=L}_{x=-L}dz}\,.$$ In Figures 1, 2, 3 and 4, a non-vanishing distribution of Burgers vectors for the regular pp-wave is displayed. They are consistent with the polarization of the chosen $H$ solution. It is worth noting that the Burgers vector has a direct relationship with the strain tensor, despite the latter coming from the metrical tensor while the first comes from the space-time torsion.  With $J = 0$ we have only the component $(3)(3)$ of the strain tensor, and at the same time there is only the component in the z direction of the Burgers vector, in all polarizations.  That is, the choice of $H$ is the only determinant for the evolution of the system.

The most interesting solution is obtained by choosing a circuit perpendicular to the axis of propagation Z. In order to present this solution, we consider a circular path, centered at the propagation axis, and focus only on the gyratonic wave. 
The Burgers vector is
\begin{equation}
b^{(i)}=\int_{0}^{2\pi}{e'^{(i)}\,_{2}d\phi}=\int_{0}^{2\pi}{e'^{(3)}\,_{2}d\phi}\,,
\end{equation}
where $e^{\prime}$ is given by (\ref{tetcil}).
Similarly to the above cases, the only non-vanishing component is
\begin{equation}
b'^{(3)}=b=-\frac{1}{\sqrt{2}}\int_{0}^{2\pi}{\frac{J}{A}d\phi}\,.
\end{equation}
We can see that the Burgers vector is zero if evaluated for a regular pp-wave in this circuit, i.e., $J=0$.
An interesting result occurs when we have an axially symmetric wave,
i.e., $J=J(u)$ and $H=H(u,\rho)$. In this particular case it is possible to evaluate the integral analytically, obtaining
\begin{equation}
b=-\frac{2\pi J}{\sqrt{2} A}\,. \label{b}
\end{equation}
We can define a dislocation core and write the strain tensor as a function of the Burgers vector. The strain tensor can be transformed into polar coordinates as
\begin{equation}
\varepsilon_{(\phi)(z)}=-\sin\phi\,\varepsilon_{(1)\left(3\right)}+\cos\phi\,\varepsilon_{(2)(3)}\,.
\end{equation}
Thus, using (\ref{3dd}) and (\ref{b}), we obtain
\begin{equation}
\varepsilon_{(\phi)(z)}=\frac{b}{2\pi\rho}\,,
\end{equation}
The result above is exactly the same result observed in a crystal with a screw dislocation, as can be seen in equation (5.3) of~\cite{Smallman}. The obtained stress field, outside the dislocation core, falls off as $\rho^{-1}$, therefore consists in a long range field. This fact can cause a particle to feel the effects of  dislocation even outside the core. 

\section{Gravitational pressure} \label{sec.5}

In this section we aim to calculate the gravitational force imparted by the gyratonic gravitational waves.  We calculate the torsion components.  The non-vanishing components are \cite{tor}
\begin{equation}\nonumber
\begin{split}
&T^{(0)(0)(1)}=-T^{(3)(1)(3)}=-\frac{1}{4\rho A^{2}}\left(2\sqrt{2}\sin{\phi}\partial_{t}J-\sin{\phi}\partial_{\phi}H+\rho\cos{\phi}\partial_{\rho}H\right)\,,\\
&T^{(0)(0)(2)}=-T^{(3)(2)(3)}=-\frac{1}{4\rho A^{2}}\left(-2\sqrt{2}\cos{\phi}\partial_{t}J+\cos{\phi}\partial_{\phi}H+\rho\sin{\phi}\partial_{\rho}H\right)\,,\\
&T^{(0)(0)(3)}=T^{(3)(0)(3)}=-\frac{1}{4A^{3}}\partial_{t}H\,,\\
&T^{(0)(1)(3)}=\frac{1}{2\rho A^{2}}\left(\sqrt{2}\sin{\phi}\partial_{t}J-\sin{\phi}\partial_{\phi}H+\rho\cos{\phi}\partial_{\rho}H\right)\,,\\
&T^{(0)(2)(3)}=\frac{1}{2\rho A^{2}}\left(-\sqrt{2}\cos{\phi}\partial_{t}J+\cos{\phi}\partial_{\phi}H+\rho\sin{\phi}\partial_{\rho}H\right)\,,\\
&T^{(3)(0)(1)}=-\frac{1}{\sqrt{2}\rho A^{2}}\sin{\phi}\partial_{t}J\,,\\
&T^{(3)(0)(2)}=\frac{1}{\sqrt{2}\rho A^{2}}\cos{\phi}\partial_{t}J\,.
\end{split}
\end{equation}
Then, using the above quantities, we have
\begin{equation}\nonumber
\begin{split}
&\Sigma^{(0)01}=\Sigma^{(3)01}=-\frac{1}{4\sqrt{2}}\frac{\partial_{\rho}H}{\sqrt{2-H}}\\
&\Sigma^{(0)02}=\Sigma^{(3)02}=-\frac{1}{4\sqrt{2}\rho^{2}}\frac{\partial_{\phi}H}{\sqrt{2-H}}+\frac{\partial_{t}J}{2\rho^{2}\sqrt{2-H}}\\
&\Sigma^{(1)01}=\frac{1}{4}\frac{\partial_{t}H}{2-H}\cos{\phi}\\
&\Sigma^{(1)02}=-\frac{1}{4\rho}\frac{\partial_{t}H}{2-H}\sin{\phi}\\
&\Sigma^{(1)03}=-\frac{1}{4\rho}\frac{\partial_{\phi}H\sin{\phi}-\rho\partial_{\rho}H\cos{\phi}}{2-H}\\
&\Sigma^{(2)01}=\frac{1}{4}\frac{\partial_{t}H}{2-H}\sin{\phi}\\
&\Sigma^{(2)02}=\frac{1}{4\rho}\frac{\partial_{t}H}{2-H}\cos{\phi}\\
&\Sigma^{(2)03}=\frac{1}{4\rho}\frac{\partial_{\phi}H\cos{\phi}+\rho\partial_{\rho}H\sin{\phi}}{2-H}\,.
\end{split}
\end{equation}
The non-null components of the energy-momentum tensor are
\begin{equation}
t^{00}=t^{03}=t^{33}=-\frac{k}{8\rho^{2}A^{2}}\left[\rho^{2}(\partial_{\rho}H)^{2}+(\partial_{\phi}H)^{2}+2\partial_{u}J\partial_{\phi}H\right]\,.
\end{equation}
Finally, the gravitational pressure defined by (\ref{19}) is given by
\begin{equation}
\phi^{(3)3}=-\frac{k}{8\rho A^{3}}\left[\rho^{2}(\partial_{\rho}H)^{2}+(\partial_{\phi}H)^{2}+2\partial_{u}J\partial_{\phi}H\right]\,.
\end{equation}
One can calculate the total force by integrating over a surface $S$,
$$
F^{(i)}=-\int{\phi^{(i)j}dS_{j}}\,.
$$
Hence, if a surface whose normal vector is oriented towards the z axis is chosen, then
\begin{equation}\label{forcez}
F_{z}\equiv F^{(3)}=\int{\phi^{(3)3}dS_{3}}=\frac{k}{8}\int d\rho d\phi \frac{\rho^{2}(\partial_{\rho}H)^{2}+(\partial_{\phi}H)^{2}+2\partial_{u}J\partial_{\phi}H}{\rho A^{3}}\,.
\end{equation}
We see that the direction of the force is longitudinal, and therefore the gravitational wave imparts  a force on hypothetical particles along the direction of the propagation of the wave.

In the case of an axially symmetric wave, i.e., $H=H(u,\rho)$ and $J=J(u)$, the integral (\ref{forcez}) can be evaluated analytically, where the surface of integration S is chosen as a disc centered in the $z$ axis, with $\rho_1$ and $\rho_2$ as the inner and outer radii, respectively. In this case, where the solution of the Einstein equation (\ref{einseq}) is given by (\ref{eq13}), we obtain
\begin{equation}
F_{z}=-\frac{C_{0}}{16}\left(\frac{1}{A(u,\rho_{2})}-\frac{1}{A(u,\rho_{1})}\right)\,,
\end{equation}
where we considered $a=1$ in the solution (\ref{eq13}).

The gravitational force can be calculated numerically for the different solutions.  In figure 5 we see the respective graphs as a function of the z coordinate, specifically at $t = 0$.  We can see that this is a negative force for all cases considered. There is a deformation parallel to the Burgers vector itself, as indicated by the strain tensor component $\varepsilon_{(3)(3)}$.

\begin{figure}[htbp]\label{fig7}
	\centering
		\includegraphics[width=0.9\textwidth]{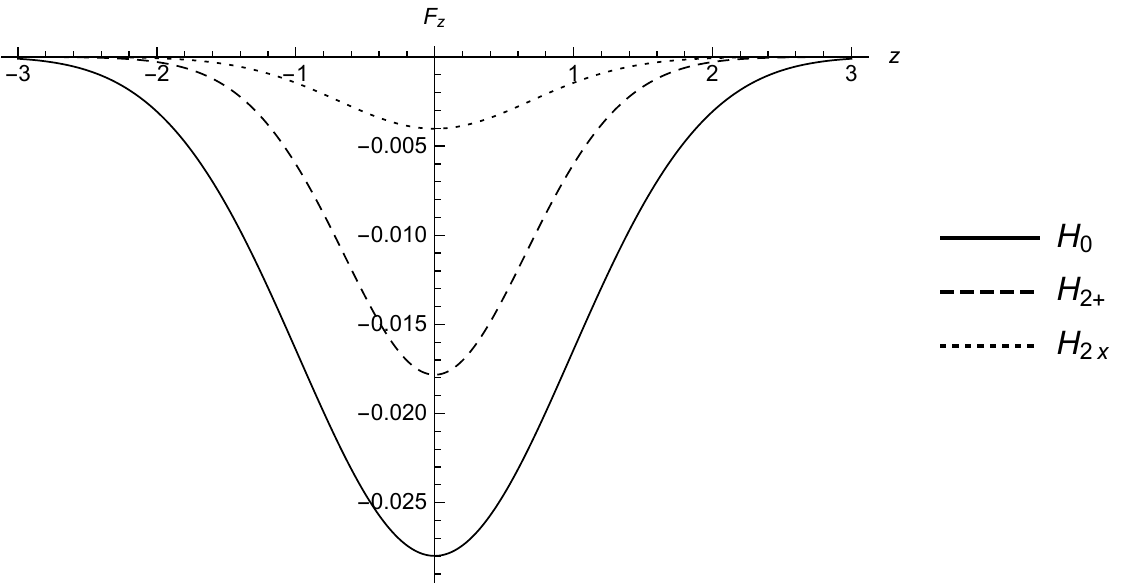}
	\caption{Forces for solutions (\ref{eq13},\ref{eq16},\ref{eq17}) with $C_{0}=1/4$, $C_{2+}=C_{2X}=1$, $a=1$, $t=0$ and $f(u)=e^{-u^{2}}=J$.}
\end{figure}

\section{Conclusion} \label{5}

In this article we analyzed how generalized pp-waves can be interpreted as topological defects.  For this purpose, we calculated the strain tensor associated with such waves, as well as the dislocation determined by the Burgers vector.  In particular, we chose a square path in the YZ and XZ planes and obtained a distribution of Burgers vectors for several $H$ solutions. With that we could see that  there is a well defined Burgers vector for regular pp-waves.  In the same sense, we chose a circular path perpendicular to the z axis that allowed us to obtain the strain tensor as a function of the modulus of the Burgers vector.  Thus we compare the result with a dislocation in a crystal.  Surprisingly, we saw that the gyratonic wave shares similarities with a crystal endowed with a topological defect with cylindrical symmetry.  We conclude that in order to describe a given metric as a topological defect, it is necessary to take into account both the strain tensor and the Burgers vector.

The results obtained with the gyratonic waves are very similar to those of crystals, mainly between the gyratonic wave and a crystal with a screw dislocation. The qualitative differences arise due to the presence of a normal component $\sigma_{(z)(z)}$ in the strain tensor of the pp-waves, while in the case of a crystal screw dislocation, only the shear component $\sigma_{(\phi)(z)}$ is present. For gravitational waves, the existence of the normal $\sigma_{(z)(z)}$ component is inherent to its type, while the existence of the shear component $\sigma_{(\phi)(z)}$ depends on the existence of the gyratonic term $J$.
Therefore, in the space-time of a gravitational wave we may have compression and shear when a longitudinal force is applied. Most interesting, when we dismiss the gyratonic term $J$ in the line element, the pp-waves space-time looses its capacity to shear.

The characterization of waves as topological defects can be applied in an attempt to quantize pp-waves. The quantization of a space-time may be performed by the geometric assumption of the Burgers vector being an integer of the Planck’s length \cite{magnon1991spin,ross1989planck} and its quantization parameters can be measured by analyzing the interaction of the gravitational field with particles \cite{quantiburgers}. Thus, interpreting pp-waves as space-time defects may provide a way to quantize these waves. For instance, in the case of an axially symmetric gravitational wave, we have $b=-\sqrt{2}\pi J/A$, thus imposing $b=nb_{0}$ \cite{quantiburgers}, where $b_{0}$ is the fundamental scale of the defect and $n$ an integer, we have $-\sqrt{2}\pi J/A=nb_{0}$. This feature will be further investigated elsewhere.


\end{document}